\documentclass[useAMS,usenatbib]{mn2e}
\usepackage[pdftex]{graphicx}

\def\spose#1{\hbox to 0pt{#1\hss}}
\def\lta{\mathrel{\spose{\lower 3pt\hbox{$\mathchar"218$}}
     \raise 2.0pt\hbox{$\mathchar"13C$}}}
\def\gta{\mathrel{\spose{\lower 3pt\hbox{$\mathchar"218$}}
     \raise 2.0pt\hbox{$\mathchar"13E$}}}

\def\bea{\begin{eqnarray}}
\def\eea{\end{eqnarray}}
\def\md{{\rm d}}
\def\DM{{\rm DM}}

\title{Dense plasma dispersion of fast radio bursts}
\author[Tuntsov]{Artem V. Tuntsov\thanks{E-mail: Artem.Tuntsov@manlyastrophysics.org}\\
$\!\!$Manly Astrophysics, 3/22 Cliff St,  Manly 2095, Australia}

\begin{document}

\date{Accepted 2014 February 26.  Received 2014 February 25;  in original form 2014 February 5}

\pagerange{\pageref{firstpage}--\pageref{lastpage}} \pubyear{2014}

\maketitle

\label{firstpage}

\begin{abstract}
Stellar coronae have been invoked to explain the apparently extragalactic dispersion measures observed in fast radio bursts. This paper demonstrates that the suggested plasma densities would lead to deviations from the standard dispersion curve that are inconsistent with the data. The problem is then turned around and higher-order dispersion terms are connected to the moments of the density distribution along the line of sight. The deviations quantified in three observed bursts are analysed and a lower limit on the maximum electron density is obtained in one case, although with considerable uncertainty. Selection effects are then discussed and shown to be non-restrictive in relation to plasma density, except at the lowest frequencies and highest temperatures.
\end{abstract}


\section{Origins of Fast radio bursts}
\label{section:intro}

Fast radio bursts (FRBs) are a recently discovered class of bright ($\sim 1\,\mathrm{Jy}$) non-repeating radio transients of $\sim1\,\mathrm{ms}$ duration \citep{lorimeretal2007, keaneetal2012, thorntonetal2013}. The bursts arrival time $t$ vs. observed frequency~$f$ plots adhere to the cold plasma dispersion law, $t\propto f^{-2}$, very accurately and high values of their dispersion measure, $\DM\sim(300-1000)\,\mathrm{cm}^{-3}\,\mathrm{pc}$ observed far from the Galactic plane suggest an extragalactic origin for FRBs at cosmological distances, $z\gta0.1$. 

The distances, considerable flux densities and short durations imply energies of the FRB sources of up to $10^{40}\,\mathrm{erg}$ and brightness temperatures of order $10^{33}\,\mathrm{K}$. This can only be achieved by coherent emission with exteme bunching ratios, $q/e\sim10^{15}$ and involving highly relativistic outflows with Lorentz factor of order $\Gamma\gta10^2$ at the lowest \citep{katz2013}. A number of candidate astrophysical phenomena have been suggested to power the bursts, including magnetar hyperflares \citep{popovpostnov2007}, supernova impacts on magnetospheres of their companion neutron stars \citep{egorovpostnov2009}, white dwarf or neutron star mergers ({Kashiyama}, {Ioka} \&
  {M{\'e}sz{\'a}ros} 2013; Pshirkov \& Postnov 2010; \citealt{totani2013}), supramassive neutron star collapse into a black hole \citep{falckerezzolla2013,zhang2014} and evaporation of primordial black holes \citep{keaneetal2012}. Recently, \cite{kulkarnietal2014} have extensively reviewed various proposals singling out the hyperflare model as the most attractive. However, despite the success of these models in predicting the rate, duration and energetics of the observed FRBs, explanations of the emission mechanism, in particular the required bunching ratios and Lorentz factors have so far, with a notable exception of \cite{lyubarsky2014}, been phenomenological or altogether {\it ad hoc}.

This has led some authors to question the extragalactic -- or, indeed, extraterrestrial -- origin of the FRB activity. An archival search of the data from the Parkes telescope prompted by \cite{lorimeretal2007} resulted in the discovery of perytons, a class of apparently terrestrial signals that share some, though not all, of the properties of the original FRB (`the Lorimer burst`), which has cast doubt on its extragalactic interpretation \citep{burkespolaoretal2011}; recently, perytons have been confirmed as a worldwide phenomenon, not limited to Parkes (Saint-Hilaire, Benz \& Monstein 2014). Nevertheless, the differences to the Lorimer burst
remained while subsequent discoveries of five additional FRBs \citep{keaneetal2012, thorntonetal2013} which diverge even more from the perytons in their properties seem to have established FRBs as a separate class of transients likely of celestial origin.

Loeb, Shvartzvald \& Maoz (2014) suggested an alternative interpretation of the high value of the dispersion measure characteristic of FRBs. The authors propose that, rather than being due to a cosmological path length, the observed column density of the electrons is primarily contributed by the high density of electrons, $n_e\sim10^{10}\,\mathrm{cm}^{-3}$ in the coronal plasma of a main-sequence star, which integrates to a $\DM\sim300\,\mathrm{cm}^{-3}\,\mathrm{pc}$ over a stellar scale path length of $10^{11}\,\mathrm{cm}$. If an FRB source is located at the base of the corona of a star within $\sim1\,\mathrm{kpc}$ of the Sun, the apparent brigthness, duration and rate of the FRBs might be consistent with the properties of rare, most powerful coherent radio bursts observed at some flaring stars. These bursts are thought to occur via the cyclotron maser mechanism and
do not require extreme physical conditions at their sources. 

\cite{luan2014} has criticised the flaring star interpretation by computing the free-free absorption in the coronal plasma. The observed DM implies the absorption that would conceal any radio signal generated below the corona unless it is unrealistically extended or hot.
For the burst to remain visible then, it needs to accumulate most of its DM beyond the corona, largely stripping the model of \cite{loebetal2014} of its explanatory power.

The present paper considers another effect of the high electron densitiy invoked by \cite{loebetal2014}. 
Section~\ref{section:dispersion} shows that it is high enough for dispersion law terms beyond the standard $f^{-2}$ to become important, and they are not observed.
Conversely, Section~\ref{section:moments} explains how deviations from the standard curve that are visible in the data could constrain the moments of the electron density distribution along the line of sight; estimates available for three FRBs are discussed. Section~\ref{section:discussion} concludes the paper by discussing the role of selection effects.

\section{Dispersion in dense plasma}
\label{section:dispersion}

The cold plasma dispersion relation expresses the magnitude of the wave vector $k$ of a plane electromagnetic wave in a plasma as a function of the wave frequency $\omega=2\pi f$ through
\bea
c^2 k^2(\omega)=\omega^2-\omega_p^2, \hspace{1.cm}\omega_p^2\equiv4\pi c^2 r_e n_e
\eea
where $r_e$ is the classical electron radius, $c$ -- the speed of light and $n_e$ -- the electron number density in the plasma. In the Born approximation, the arrival of a signal travelling through the plasma with group velocity $\md\omega/\md k$ along the line of sight would be delayed by ($f_p\equiv\omega_p/2\pi$)
\bea\label{taun}
\tau=\int\md D\,\left(\frac{\md\omega}{\md k}-\frac1c\right)=\int\frac{\md D}c\,\left[\left(1-\frac{f_p^2}{f^2}\right)^{-1/2}-1\right]
\eea
with respect to the (not directly observable) arrival time $t_0$ had the signal travelled through vacuum. The usual dispersion law is obtained by expanding the integrand in $f_p^2/f^2$ and only retaining the leading term:
\bea\label{tauseries}
t-t_0\simeq\int\frac{\md D}c\,\left[\frac12\frac{f_p^2}{f^2}+\frac38\left(\frac{f_p^2}{f^2}\right)^2+\frac5{16}\left(\frac{f_p^2}{f^2}\right)^3+...\right]\\\nonumber
\approx\frac{c r_e}{2\pi f^2}\DM, \hspace{.7cm}\DM\equiv\int n_e(D)\, \md D.\hspace{1.5cm}
\eea
This tuncation is not justified, however, when the electron density becomes comparable to the transparency limit
\bea\label{nfdef}
n_f=\frac{\pi f^2}{c^2 r_e}\approx1.24\times10^{10}\,\mathrm{cm}^{-3}\left(\frac{f}{\mathrm{GHz}}\right)^2
\eea
and even at $n\ll n_f$, truncation biases the inferred DM high, which can be seen by observing that all higher-order terms in the expansion~(\ref{tauseries}) are positive. The High Time Resolution Universe (HTRU) survey (Keith et al. 2010) used in~\cite{thorntonetal2013} observes at frequencies down to $f_0=1.182\,\mathrm{GHz}$ resulting in $n_0\approx1.733\times10^{10}\,\mathrm{cm}^{-3}$ -- {\it i.e.} of the order $n_e\sim10^{10}\,\mathrm{cm}^{-3}$ suggested in~\cite{loebetal2014} to explain the DM of observed FRBs. 

It is useful to specify a simple model of a uniform plasma blanket with a constant electron number density $n_e(D)=n_u$ over a depth of $D_u$ -- the same as used by~\cite{luan2014}. In the low density approximation, $n_u\to0$, the two parameters are degenerate and the pulse arrival time difference at a pair of frequencies, $f_\mathrm{lo}, f_\mathrm{hi}$ is given by the product $\DM=n_u D_u$:
\bea\label{lineariseddc}
\Delta t=t_\mathrm{lo}-t_\mathrm{hi}=\frac{c r_e}{2\pi}\DM\left(f_\mathrm{lo}^{-2}-f_\mathrm{hi}^{-2}\right)
\eea
However, for denser plasmas $n_u$ and $D_u$ decouple:
\bea\label{uniformdc}
\Delta t=\frac{D_u}{c}\left[\left(1-\frac{c^2 r_e n_u}{\pi f_\mathrm{lo}^2}\right)^{-1/2}-\left(1-\frac{c^2 r_e n_u}{\pi f_\mathrm{hi}
^2}\right)^{-1/2}\right]
\eea
allowing one to constrain both if the data permit. More intricate two-parameter distributions of the electron density along the line of sight -- {\it e.g.}, power-law models of stellar coronae -- can be easily mapped onto this minimal model.


\begin{figure}
\includegraphics[height=43mm]{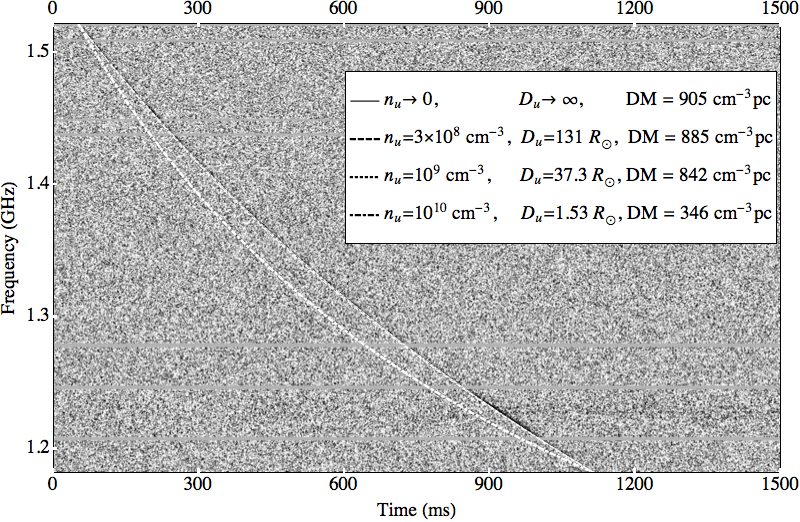}
\includegraphics[height=43mm]{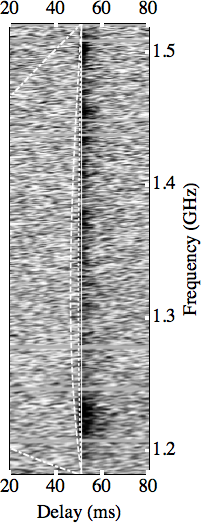}
\vskip-0.2truecm
\caption{{\it Left:} The dynamic spectrum of FRB110220 \citep{thorntonetal2013} along with the four model dispersion curves of varying electron density. {\it Right:} The same dynamic spectrum transformed by advancing the time series, at each separate frequency, by the delay predicted by the best-fit low-density model. The same transform is applied to all dispersion curves. Only the fragment centred on the pulse is shown; please note a different horizontal scale.}
\label{figure:dcurves}
\medskip
\end{figure}

Figure~\ref{figure:dcurves} compares the dispersion curves~(\ref{uniformdc}) for uniform models of various densities $n_u$ including the low-density limit~(\ref{lineariseddc}) to the behaviour of the highest signal-to-noise ratio FRB observed so far, FRB110220, displayed in~\cite{thorntonetal2013}. The values of $D_u$ ($\DM$ in the low-density limit) at each value of $n_u$ are obtained by applying~(\ref{lineariseddc},\ref{uniformdc}) to the arrival time moments read off the top and bottom of the plot\footnote{The signal-to-noise ratio for these frequencies is generally lower due to the instrumental roll-off. This might explain somewhat (four per cent) lower estimate of the DM with~(\ref{lineariseddc}) than the value reported by~\cite{thorntonetal2013} obtained by fitting the dispersion curve measured along the entire bandpass. This said, relaxing the position of pivot points along the dispersion curve did not result in a significantly closer match.}. 

Altough an illustration only, the figure clearly rules out the extreme values of the electron density $n_u\sim10^{10}\,\mathrm{cm}^{-3}$. The respective line posseses too much curvature compared to the data even though its formal $\DM$ is $\sim350\,\mathrm{cm}^{-3}\mathrm{pc}$ only, nearly three times as low as the estimate for the low-density model; this discrepancy cannot be reduced much by choosing different pivot points $f_\mathrm{lo}, f_\mathrm{hi}$ along the observed dispersion curve. The $n_u=10^9\,\mathrm{cm}^{-3}$ appears to perform reasonably well at first but a closer look reveals that it is not consistent with the data either. The right panel of the figure displays the dynamic spectrum `incoherently dedispersed' with the obviously best-performing $n_u\to0$ model along with the four dispersion curves dedispersed in the same fashion. The panel testifies to the quality of the low-density dispersion curve (now a straight vertical line) in fitting the data and clearly rules out the $n_u=10^9\,\mathrm{cm}^{-3}$ curve. Only $n_u\lta10^8\,\mathrm{cm}^{-3}$ curves seem to be permitted but even the $n_u=3\times10^8\,\mathrm{cm}^{-3}$ curve, which is only allowed marginally, requires the uniform plasma depth $D_u>100\,\mathrm{R}_\odot$. This makes the main-sequence star origin of FRB110220 less plausible. 

Analysis of two other bursts with published dynamic spectra leads to similar conclusions, albeit somewhat less restrictive:  $n_u\lta1.5\times10^9\,\mathrm{cm}^{-3}$, $D_u\gta20\mathrm{R}_\odot$ for FRB010621 and $n_u\lta2\times10^9\,\mathrm{cm}^{-3}$, $D_u\gta7\mathrm{R}_\odot$ for FRB010724.

\section{Charting electron distribution}
\label{section:moments}

It is now convenient to reference the electron density to the transparency limit~(\ref{nfdef}) at the lowest observed frequency $n_0\equiv n_f(f_0)$.
Waves below the plasma frequency reflect off or decay exponentially in the plasma and do not reach the observer; therefore, for observed pulses $n_e<n_0$ anywhere along the line of sight. Using $n_0$, (\ref{tauseries}) can be rewritten:
\bea\label{tf}
t=t_0+a_1\frac{f_0^2}{f^2}+a_2\left(\frac{f_0^2}{f^2}\right)^2+a_3\left(\frac{f_0^2}{f^2}\right)^3+...
\eea
with coefficients at successive powers of $f_0^2/f^2$ proportional to the moments of $n_e/n_0$ along the line of sight
\bea
a_k=\frac{(2k-1)!!}{(2k)!!}\int\frac{\md D}{c}\,\left(\frac{n_e}{n_0}\right)^k,\,\mathrm{~~including~}a_1=\frac\DM{2cn_0}.
\eea
They have the dimension of time and given wide bands of modern radio surveys
measure the contribution of the $a_k (f_0^2/f^2)^k$ term to the pulse sweep across the bandpass.

For a uniform plasma model the amplitudes are simply
\bea\label{aku}
a_k=\frac{(2k-1)!!}{(2k)!!}\frac{D_u n_u^k}{c n_0^k}
\eea
whereas for a stellar corona with a power law distribution of the electron density $n_e(R)=n_p(R_p/R)^p,\,R>R_p$ they are, assuming the index $p>1$ and integration along a radial line of sight to the base at $R_p$ from the stellar centre,
\bea\label{akp}
a_k=\frac{(2k-1)!!}{(2k)!!}\frac1{kp-1}\frac{R_p n_p^k}{c n_0^k}.
\eea
One can notice that, as a function of $k$, (\ref{aku}) decays slower than~(\ref{akp}). In fact, 
for an arbitrary distribution $n_e(D)$ on a transparent line of sight the coefficients 
respect the following inequality hierarchy:
\bea\label{hierarchy}
a_{k+1}\leq\frac{2k+1}{2k+2}\frac{n_m}{n_0} a_k,  \hspace{.5cm}\mathrm{where~} n_m\equiv\max n_e(D)
\eea
with equality only attained for~(\ref{aku}), with $n_e(D)\in\{0,n_u\}$. 

For instance, $n_u\lta3\times10^8\,\mathrm{cm}^{-3}\approx0.02n_0$ obtained above for FRB110220 implies $a_2<0.02 a_1$ and the contribution of the $f^{-4}$ term to the arrival time difference at the ends of the bandpass is no more than $20\,\mathrm{ms}$ (most of which  is absorbed into the bias, cf.~(\ref{abias})). 
The amplitudes of the third and further terms
are below the smearing timescale of $\sim2a_1/N$ of a sweeping pulse due to a non-zero width of each of $N\sim10^3$ channels. Therefore, if FRB110220 is representative, we do not expect modern surveys to be sensitive to terms beyond $f^{-4}$. Up to this accuracy, any power-law corona may be represented by an `effective' uniform plasma blanket, obtained by equating $a_{1,2}$ of the two models:
\bea
n_u^\mathrm{eff}=n_p\frac{p-1}{2p-1}, \hspace{.5cm} D_u^\mathrm{eff}=R_p\frac{2p-1}{(p-1)^2}.
\eea

For a general distribution, constraining $n_m$ from above is not straightforward and thus estimating $a_k$ amplitudes {\it a~priori} is not possible. Instead, when the presence of $f^{-4}$ contribution to the dispersion curve has been reliably established (and attributed to plasma dispersion
), inequality~(\ref{hierarchy}) with $k=1$ can be used to place a \emph{lower} limit on the maximum electron density reached along the line of sight:
\bea\label{nm}
n_m \gta\frac{4\pi f_0^2}{3c^2r_e}\frac{a_2}{a_1}
\eea
allowing to probe densities below the transparency limit~(\ref{nfdef}).

Estimating the coefficients $a_k$ requires accurate measurement of the pulse arrival time at a range of frequencies and fitting the resulting dispersion curve with a polynomial in $f^{-2}$. At present it is not customary to report the results of such fits and the author is not aware if they have been attempted. The current practice is to report, where the data permit, the degree to which the index $\alpha$ in the $t\propto f^\alpha$ fit to the data is consistent with the standard value of $-2$.

It is possible to relate the parameters of the polynomial and power-law fits statistically by correlating the model predictions for observables used to estimate $\alpha$, although the correlation would depend on the detail of the estimation procedure that are not readily available in the literature. However, given that both fitting models have a common, low-density, limit at $a_{k\geq2}\to0$ and $\alpha\to-2$, and that the reported values of $\alpha$ are at worst only marginally inconsistent with the limit, it might be appropriate to use the two models' relation to a few common statistics as a proxy for the correlation analysis. If terms higher than $f^{-4}$ in the polynomial fit are neglected, the averages of the first three terms of the Taylor series in $f^{-2}$ are sufficient for matching:
\bea\nonumber
t=\hat{t}_0+\hat{a}x^{-\alpha/2}\,\leftrightarrow\, t=t_0+a_1 x+a_2x^2, \hspace{.3cm}x\equiv\left(f_0/f\right)^2\\
\nonumber\hat t_0=t_0+a_2\langle x^2\rangle\hspace{1.6cm} t_0=\hat t_0+\frac{\alpha+2}2\frac{\hat a\langle x^2\rangle}{2\langle x\rangle}\hspace{.9cm}\\
\label{mapping}\hat{a}=a_1-2a_2\langle x\rangle\langle\log x\rangle\hspace{.1cm}\Leftrightarrow\hspace{.1cm} a_1=\hat a\left(1-\frac{\alpha+2}2\langle\log x\rangle\right)\hspace{.2cm}\\
\nonumber\alpha=-2-4\frac{a_2}{a_1}\langle x\rangle\hspace{1.4cm}a_2=-\frac{\alpha+2}2\frac{\hat a}{2\langle x\rangle}\hspace{1.4cm}
\eea
where $\langle\cdot\rangle$ are (possibly, weighted) averages over the sampling points $f_i$. For unweighted datasets of \cite{thorntonetal2013} and \cite{keaneetal2012}, respectively, $(\langle x\rangle, \langle x^2\rangle, \langle\log x\rangle)$ are $(0.777, 0.616, -0.263)$ and $(0.810, 0.667, -0.218)$.

\begin{table}
\caption{Estimates of the dispersion curve power index $\alpha$ available in the literature (010621 -- \citealt{keaneetal2012}; 110220, 110703 -- \citealt{thorntonetal2013}); their $\DM$ values are also quoted. The next four rows show the true (\ref{mapping}) and apparent (\ref{deltat}) deviations of the dispersion curves from the low-density limit \emph{assuming} $\alpha$ deviation is due to density correction, as well as the density $n_u=4n_0a_2/3a_1$ and depth $D_u=3ca_1^2/a_2$ of the effective uniform model.}
\label{table:deltaestimates}
\centering
\begin{tabular}{cccc}
\hline
FRB & 010621 & 110220 & 110703\cr
\hline
$(\alpha+2)\times10^3$ & $-20\pm10$ &  $-3\pm6$ & $0\pm6$ \cr
$\DM (\mathrm{cm}^{-3}\mathrm{pc})$ & $746\pm1$ & $944.38\pm0.1$ & $1103.6\pm0.7$ \cr
$a_2(\mathrm{ms})$ &  $13\pm6$ & $3\pm5$ &$0\pm6$\cr
$||\delta\hat t|| (\mathrm{ms})$ &  $0.5\pm0.2$ &  $0.1\pm0.3$ &  $0\pm0.3$ \cr
$n_u (10^7\,\mathrm{cm}^{-3})$ &$15\pm7$& $2\pm4$& $0\pm4$\cr
$D_u (\mathrm{R}_\odot)$ &$200\pm100$& $>600$& $>1100$\cr
\hline
\end{tabular}
\end{table}

As both amplitudes $a_{1,2}$ are positive, we expect $\alpha<-2$. Interestingly, of the three available estimates of the FRB dispersion curve power-law index, none is suggestive of $\alpha>-2$ while two actually favour $\alpha<-2$, even though with considerable uncertainty. Table~\ref{table:deltaestimates} presents the details of these FRBs along with parameters of the dense plasma model that would correspond to the measured value of $\alpha$. One of the bursts, FRB010621, shows an indication of extra curvature at the level of $2\sigma$. It might be interesting to check if a polynomial in $f^{-2}$ fits the data much better than the power law. 
If so, $n_e$ is expected to reach $n_m\approx(1.5\pm0.7)\times10^8\,\mathrm{cm}^{-3}$.

It is worth stressing, however, that a significant $a_2$ does not necessarily imply a deviation of this scale in the dynamic spectrum dedispersed with the best-fit low-density model, because the parameters of the latter would be biased. Forcing $\alpha=-2$ in the derivation similar to that leading to~(\ref{mapping}) results in
\bea\label{abias}
\alpha\equiv-2:\hspace{.1cm}\hat t_0=t_0+a_2\left(\langle x^2\rangle - 2\langle x\rangle^2\right),\, \hat a=a_1+2 a_2\langle x\rangle
\eea
hence the true and best-fit low-density models differ by
\bea
\delta\hat t\equiv\left.\hat{t}\right|_{\alpha\equiv-2}-t=a_2\left[\langle x^2\rangle-\langle x\rangle^2-\left(x-\langle x\rangle\right)^2\right],
\eea
which 
has a variation of
\bea\label{deltat}
||\delta\hat t||\equiv\max\delta\hat t-\min\delta\hat t=a_2\max\limits_x\,\left(x-\langle x\rangle\right)^2
\eea
equal to $a_2(1-\langle x\rangle)^2$, or only $0.05a_2$ and $0.036a_2$ for~\cite{thorntonetal2013} and~\cite{keaneetal2012}, respectively. These values, also quoted in the table, are below the temporal resolution of both surveys and thus a polynomial re-analysis is not expected to show any advantage over the power-law fit.

Finally, it is possible to have a {\it bona fide} $\propto f^{-4}$ term without resorting to dense plasmas and relying instead on the phenomenon responsible for pulse broadening -- although through scattering on a large-scale gradient rather than stochastic fluctuations of the electron density. On a ray deflected by an angle $\theta\ll1$ the geometric time delay with respect to the unperturbed ray is $t_g=D_\mathrm{eff}\theta^2/2c$ with $D_\mathrm{eff}$ close to the smallest of the source-deflector-observer distances. The deflection angle due to a fixed transverse gradient of the electron column density is $\theta=c^2 r_e\nabla_\perp\DM/2\pi f^2$ and if this gradient is maintained over lengths $\sim\theta D_\mathrm{eff}$, the delay will scale as $t_g= a_g (f_0/f)^4$ with amplitude $a_g=D_\mathrm{eff}\theta_0^2/2c$. For $a_g=10\,\mathrm{ms}$ and a cosmological distance, $D_\mathrm{eff}=1\,\mathrm{Gpc}$, the required gradient is $\nabla_\perp\DM\approx20\,\mathrm{cm}^{-3}\mathrm{pc}_\parallel\mathrm{pc}_\perp^{-1}$ at scales $\mathrm{D}_\mathrm{eff}\theta_0\approx0.5\,\mathrm{pc}$. This would correspond to a rate $\md\DM/\md t\approx0.01\,\mathrm{cm}^{-3}\mathrm{pc}\cdot\mathrm{yr}^{-1}$ for transverse velocities $\sim10^3\,\mathrm{km}\cdot\mathrm{s}^{-1}$, which is not unreasonable in the Galaxy \citep{hobbsetal2004} although it is not clear whether the comparison is appropriate as FRB scattering is observed to be anomalously low \citep{lorimeretal2013}; in addition, finer scale $\DM$ fluctuations would need to be suppressed for a clean $\propto f^{-4}$ signal. In the Galaxy, $D_\mathrm{eff}=1\,\mathrm{kpc}$, the gradient would need to be a thousand times as high at scales a thousand times as small -- i.e., the same variation of $10\,\mathrm{cm}^{-3}\mathrm{pc}$ over a transverse separation of just $100\,\mathrm{AU}$. Importantly, any such scattering contribution \emph{is of the same sign} as the high-density deviations of the dispersion curve; therefore, scattering can only strengthen the constraints of Section~\ref{section:dispersion}.

\section{Summary. Effects of selection}
\label{section:discussion}

In Section~\ref{section:dispersion} the electron densities $n_e\sim10^{10}\,\mathrm{cm}^{-3}$ suggested by \cite{loebetal2014} to explain the dispersion measures of FRBs were shown to be inconsistent with the upper limit on the deviation of the observed dispersion curves from the standard, low-density cold plasma law. Densities that are allowed by the data, $n_e\sim10^{8-9}\,\mathrm{cm}^{-3}$, are not extreme on their own -- such is the density at the base of the corona of the Sun -- but require extreme paths, $D\sim10^2\,\mathrm{R}_\odot$, to integrate to observed DMs, bringing the validity of the stellar model into question. 

Section~\ref{section:moments} turned the problem around and introduced a~simple, moment-based framework with which the deviations of the dispersion law from its low-density limit can be analysed. It was shown that the leading term of the deviation, proportional to the second moment of $n_e$, disguises itself by biasing the estimate of the first moment, the DM, high; this leaves only 4 to 5 per cent of the signal in a form that cannot be so masked. The current surveys are unlikely to probe beyond the second moment and the data available so far are consistent with the standard law to at least $2\sigma$.

The free-free absorption argument of \cite{luan2014} 
can, in the isothermal approximation, also be reduced to a limit on the second moment of $n_e$, the emission measure: requiring the free-free optical depth $\int\md D\,n_e^2\bar\alpha_\mathrm{ff}(\Theta)$, with $\bar\alpha_\mathrm{ff}$ being the absorption coefficient of hydrogen plasma at unit density, to be below a reasonable value $\kappa\sim1$ implies
\bea\label{luanlimit}
a_2<\frac{3^{5/2}c\kappa}{2^{11/2}\pi^{5/2}r_ef_0^2}\frac{\Theta^{3/2}}{\tilde{g}_\mathrm{ff}(\Theta)}
\eea
where $\tilde{g}_\mathrm{ff}\sim10$ is a slowly-varying Gaunt factor, $\Theta=k_BT/m_ec^2$ -- temperature in units of electron rest energy and $\kappa$ the allowed optical depth. 

The limit set by the allowed deviation $||\delta\hat t||$ from the low-density dispersion curve
\bea\label{climit}
a_2<\frac{||\delta\hat t||}{\max\limits_x\, (x-\langle x\rangle)^2}
\eea
is independent of $\Theta$ and $f_0$ but~(\ref{luanlimit}) remains superior at temperatures up to $1.5\times10^6\,\mathrm{K}$, assuming $\kappa=1$, \cite{thorntonetal2013} frequency coverage and sensitivity to  deviations at $||\delta\hat t||=3\,\mathrm{ms}$ level -- five times the resolution in Figure~\ref{figure:dcurves}.

This might be a reason why no deviations from the standard dispersion curves have been reliably detected -- and ease some worries regarding FRB candidate selections. Concerns have been raised that selection criteria used might be effectively imposing cold plasma dispersion
 -- or indeed, extra-terrestrial origin -- 
on FRBs by only selecting candidates that conform to it from potentially a much broader population that does not. While some sort of selection is inevitable in the presence of noise and interference, the scope adopted might be too narrow for a new class of objects of as yet unknown origin, it has been argued. Superiority of~(\ref{luanlimit}) over~(\ref{climit}) implies that selection is not restrictive in relation to plasma density; if the signal emerges from the plasma unobsorbed, it would have the dispersion curve that does not deviate from the low-density limit at the resolution of current surveys. However, this will not necessarily be the case at lower frequencies or when the time resolution is improved. At $100\,\mathrm{MHz}$ the non-standard dispersion curve might show up for plasmas as cold as $T\sim10^4\,\mathrm{K}$ (cf. note added in proof).

More speculatively, one might extend the reasoning above to the idea of FRB stellar origin itself. Could they originate in a relatively frequent phenomenon in dense stellar envelopes, of which FRBs is a fraction that happens at a density sufficiently low to let the FRB make it to the observer?\footnote{\cite{kulkarnietal2014} acknowledge, parenthetically, similar possibility  in relation to the absorption within FRB host galaxies.} If so, the estimates given above suggest the envelopes have to be rather hefty, extending to tens and hundreds solar radii. And although coronae and ionised outflows of these scales are not out of the question, they are certainly less numerous than the main sequence or flaring stars.

\section*{Acknowledgments}
The author thanks the participants of The Ephemeral Universe meeting in Perth for a lively discussion of FRB candidate selection, the referee for pointing out how scattering can mimic the dense plasma signal, the editor for their advice on the presentation of the paper and Mark Walker for valuable suggestions. 

\section*{Note added in press}
After this Letter has been accepted, Prof. Loeb kindly made the author aware of a population of flares in pre-main sequence stars with characteristic scales and temperatures of up to $100\,\mathrm{R}_\odot$ and $10^9\,\mathrm{K}$. Rather narrowly, the most extreme of such flares evade the constraints presented in this paper and still rarer events might be contemplated that do so more comfortably. Importantly, the free-free absorption is much suppressed at these temperatures and deviations from the low-density dispersion law can be detected. A separate study is needed to see if the lack of  deviations in the present data can be due to selection. If it is, the effect described in this paper can, instead of ruling out, confirm the model of Loeb et al. (2014).

\bibliography{dc}{}
\bibliographystyle{mn2e}

\label{lastpage}
\end{document}